# Universal scaling of resistivity in bilayer graphene

Kalon Gopinadhan, Young Jun Shin, and Hyunsoo Yang[a)]

*Department of Electrical and Computer Engineering, NUSNNI-Nanocore, Graphene Research Centre, National University of Singapore, 117576 Singapore*

We report the temperature dependent electrical transport properties of gated bilayer graphene devices. We see a clear evidence of insulating behavior due to electron-hole charge puddles. The electrical resistivity increases while the mobility decreases with decreasing temperature, a characteristic due to carrier inhomogeneity in graphene. The theoretical fittings using an empirical formula of single electron tunneling indicate that electrical resistivity follows a universal curve with a scaling parameter. The scaling parameter is determined to be a measure of the fluctuations in the electron-hole puddle distribution.

[a)]E-mail address: eleyang@nus.edu.sg

Graphene is the thinnest electronic material in two dimensions, in which a single layer of carbon atoms are arranged at the six corners of a hexagonal honey-comb lattice structure. It is electronically semi-metallic with a linear band structure at low energies and hence a zero electron (hole) effective mass.[1-3] In addition, a very weak scattering from phonons implies a high mobility for the carriers. For example, a very large mobility of 200,000 $cm^2V^{-1}s^{-1}$ at 4 K is reported in suspended single layer graphene.[4, 5] The electric field tunability of its ambipolar characteristic makes graphene very unique from other two dimensional materials. Even though graphene possesses many unique properties in comparison to Si, the absence of an electronic band gap in single layer graphene is one of drawbacks for transistor (digital) applications, however it can be used in analogue, high frequency applications. On the other hand, bilayer graphene may be more suitable both for digital and analogue applications, since a band gap can be induced by the application of a gate voltage.[6] The spin-orbit interaction is also not negligible in bilayer graphene in contrast to single layer graphene, which may lead to gate tunable spin field effect transistors.[7]

Bilayer graphene is semi-metallic with a parabolic band diagram at low energies for both electrons and holes. However, reported electron transport properties indicate that bilayer graphene behaves either as an insulator where the resistivity increases with decreasing temperature or a mixture of metallic and insulating states.[8, 9] Our previous study suggests that bilayer graphene can exhibit metal insulator transitions due to a change in the distribution of electron hole puddles.[10] Several theoretical papers address this anomalous increase in resistivity and especially Hwang *et al.* tried to quantify based on thermal excitations of the carriers and carrier inhomogeneity.[11] However, different approximations are required to fit the results at low and high temperatures. There is no unified formula which can explain the electrical properties in the whole range of temperatures. In addition, it is important to understand gated bilayer graphene transistors for applications. However, when an additional parameter like the gate voltage is involved, the complexity increases. On the other hand, the electrical property of single layer graphene at different gate voltages is explained with a single formula.[12]

In this letter, we show the electrical transport properties of top and back gate bilayer graphene devices. The resistivity shows an anomalous increase with decreasing temperature, a characteristic of insulating transport. The observed resistivity variation with temperature is gate voltage tunable with a characteristic scaling parameter. The normalized resistivity follows the universal curve with a single scaling parameter, and the scaling parameter is a measure of the carrier inhomogeneity.

The graphene is prepared by micromechanical exfoliation of Kish graphite followed by a transfer to a heavily p-doped Si substrate, which has a layer of 300 nm thick $SiO_2$. The resultant graphene is identified by an optical microscope and further examined by a Raman spectrophotometer.[10, 13-16] Electrodes are fabricated by optical lithography followed by the deposition of Cr (5 nm)/Au (150 nm) using a thermal evaporator. Standard lift-off procedures using warm acetone are utilized after the deposition. For the top gate fabrication, similar procedures are used and a 10 nm of Al is deposited in two steps followed by natural oxidation. The complete oxidation of Al is confirmed by x-ray photoelectron spectroscopy. [15]

For Raman spectroscopy measurements a laser wavelength of 488 nm with a power density ~ 0.5 mW/cm$^2$ is used to avoid any laser induced heating. A Raman spectrum of pristine graphene is shown in Fig. 1(a). The prominent modes are the G mode at 1582 cm$^{-1}$, G* mode at 2454 cm$^{-1}$, and the 2D mode at 2710 cm$^{-1}$. A Lorentian peak fitting is performed on the 2D peak and a good fitting is obtained with four Lorentians. The intensity of three Lorentians is much higher than the fourth one as shown in Fig. 1(b). This implies that graphene is of bilayer with AB Bernal stacking.[17] It is known that the intensity of the Lorentians may vary depending on the laser wavelength.[18] The absence of disorder (D) peak at ~1360 cm$^{-1}$ in the Raman spectrum suggests that the absence of a significant number of defects in the sample. Even though there is no detectable defect in the sample, any carrier inhomogeneity cannot be detected by Raman spectroscopy, whereas electrical transport measurements are very sensitive to this and we do see a strong evidence for the existence of inhomogeneous charge distribution as discussed later.

The optical microscopy image of the fabricated top gated device is shown in Fig. 2(a). In order to see whether the sample exhibits ambipolar characteristics and to determine the charge neutrality point (CNP) corresponding to the intersection of the Dirac cones, the channel (source-drain) resistance is measured as a function of the top gate voltage $V_G$. Figure 2(b) shows the resistance $R_{xx}$ as a function of $V_G$ at various temperatures from 5 to 220 K. All the measurements are carried out under high vacuum conditions (< 1×10$^{-7}$ Torr) in a Helium cryostat. It is clear from Fig. 2(b) that CNP is shifted slightly to the positive side at all temperatures and the detailed measurements show that its exact value depends on the level of the applied gate voltage.[13] For gate voltages higher than the CNP, graphene is electron doped, whereas for voltages lower than the CNP it is hole doped, thereby exhibiting ambipolar characteristics. This suggests that in the absence of gate voltage, the transport of our sample is hole dominant.

In order to determine the carrier concentration in the sample, the capacitance of the gate dielectric is utilized. For an $Al_2O_3$ thickness ($d$) of 10 nm and dielectric constant ($\kappa$) of 8, the gate capacitance per unit area, $C_g = \varepsilon_0 \kappa / d$ is 7.08×10$^{-7}$ F/cm$^2$. The carrier concentration is estimated using the relation, $n = -C_g(V_G - V_D)/e$, where $V_D$ is the voltage corresponding to the CNP. At zero gate voltage, the carrier concentration shows a slight increase with decreasing the temperature from 220 to 5 K. According to Drude's theory, away from the CNP the conductivity is given by $\sigma = ne\mu$, where $\mu$ is mobility. This implies that mobility decreases with decreasing temperature from 220 to 5 K. From Fig. 2(b), it is clear that this is true for all gate voltages.

It is widely accepted that phonons play a minor role in the temperature dependent mobility of bilayer graphene, since the electric field of the substrate surface phonons is screened.[19, 20] The other possible scattering mechanisms are Coulomb scattering by charged impurities at the $SiO_2$ surface, which is responsible for the formation of electron-hole puddles, and short-range scattering by defects. The decrease in mobility with decreasing temperature suggests that the transport in bilayer graphene is determined by Coulomb scattering as reported previously.[20] It is evident from Fig. 2(b) that there appears an asymmetry in electron and hole transport. The asymmetry increases as the temperature decreases from 220 K. It is reported that in bilayer graphene, the asymmetry can appear due to several reasons related to intrinsic properties such as inter-layer coupling, next nearest neighbor hoping within each graphene plane, energy difference between dimer and non-dimer sites in a unit cell, and non-orthogonal atomic orbitals.[21] According to the reported result, this intrinsic asymmetry should be independent of temperature[19] suggesting that extrinsic factors like electron-hole puddles are responsible for the observed behavior. The increase in asymmetry with relatively small change in resistivity in the hole dominant regime ($V_G < V_D$) with decreasing temperature implies that more hole rich regions are formed in comparison to electron regions. The spatial inhomogeneity of electron and hole like regions forms a quantum dot like situation with tunneling characteristics in the disordered bilayer graphene sample. It is also noticeable in Fig. 2(b) that at low temperatures $R_{xx}$ exhibits oscillations which vanish as the temperature increases beyond 20 K. These oscillations have the characteristic of universal conductance fluctuations (UCF) and may be arising due to a possible perpendicular pseudo magnetic field in graphene as a result of ripples and the finite size of the graphene.[22] The absence of oscillations beyond 20 K suggests that a large phase coherence length of the carriers at low temperature is responsible for this characteristic.[23]

In order to rule out any effect from the contact resistance in the two-terminal measurement as well as from the trap states in the top gate geometry, we have measured the four probe resistivity from a back gated Hall bar bilayer sample. The optical image of the fabricated Hall bar is shown in Fig. 2(c). Figure 2(d) shows the resistivity as a function of the back gate voltage ($V_{BG}$) at different temperatures from 220 to 60 K. The charge neutrality point is ~ 35 V suggesting that bilayer graphene is hole doped at zero gate voltage. It is inferred from the charge neutrality point at different temperatures that there is a very small change in the carrier concentration at different temperatures. This implies that mobility decreases with decreasing temperature around the CNP, which is identical to that observed from the top gated sample, suggesting electron-hole puddles as the scattering centers.

Figure 3(a) shows the $R_{xx}$ of the top gated sample as a function of temperature ($T$) from 5 to 220 K at different $V_G$. For a particular gate voltage, the resistance increases with decreasing temperature and saturates at low temperatures, showing a characteristic of an insulator. The spatial variation of electron like and hole like regions separated by tunnel barriers forms a quantum dot like situation in the disordered bilayer graphene sample. The temperature dependence of the disordered bilayer graphene has the characteristics of single electron Kondo tunneling. Therefore, we have utilized a modified empirical formula typically used to describe single electron tunneling to fit our data,

$$R(T,V_G) = R_0(V_G) + \frac{R_e(V_G)}{((1+22.35(T/T_s(V_G))^2)^{0.22}} \qquad (1)$$

where $R_0$ is the residual resistance due to static defects present in the sample, $R_e$ is the saturated resistance, and $T_s$ is the scaling parameter which is a measure of the fluctuations of the electron-hole puddle potentials. $R_0$ and $R_e$ are the temperature independent parameters used in the fittings.[24-26] It is clear from Fig. 3(a) that Eq. (1) is a good fit to the experimental curves at all applied gate voltages including CNP. Figure 3(b) shows the resistivity of the back gated sample as a function of temperature at different back gate voltages. At all gate voltages, the resistivity shows an insulating characteristic similar to the top gated graphene sample. The resistivity data are fitted with Eq. (1) which yields a good fit. Figures 3(c) and 3(d) show the normalized resistivity (with respect to saturated resistivity) versus normalized temperature (with respect to $T_S$) at different gate voltages from the top and bottom gated sample, respectively. To our surprise, all of the curves collapse into a single curve with a characteristic scaling parameter $T_S$, suggesting the universal character of insulating bilayer graphene. This suggests that Eq. (1) can be used for characterizing insulating bilayer graphene samples for both top and back gated schemes. Similar temperature dependence data from He-ion irradiated single layer graphene was reported by other group[27], which was attributed to the Kondo effect with a large negative magnetoresistance in their samples. However, in our samples we rather see a positive magnetoresistance (see supplementary materials) due to carrier inhomogeneity thereby ruling out the possibility of a single impurity Kondo effect. The low temperature saturation in the resistivity is attributed to finite size effects.[28]

Figure 4(a) shows the variation of the derived scaling parameter $T_S$ as a function of $V_G$. The scaling parameter shows a maximum of ~15 meV at CNP where $V_G = 0.5$ V and drops on either side of the CNP. The scaling parameter is tunable from 10 to 15 meV upon applying a maximum $V_G$ of ±3 V. This also suggests that at the Dirac point, the potential fluctuation due to electron-hole puddles reaches a maximum. Figure 4(b-d) shows the variation of $R_0$, $R_e$, and $G_0$(=1/$R_0$) as a function of $V_G$, respectively. The corresponding peak or dip around the Dirac point indicates the tunability by the gate voltage and suggests the validity of the derived parameters.

Figure 5(a) shows the experimentally determined conductance as a function of temperature at the CNP from the top gated sample. The conductance of bilayer graphene is numerically estimated with different approximations as detailed in a recent report.[11] At the CNP, both electrons and holes are equally populated, and the total conductance $G_t$ due to electron and hole puddles at low temperatures ($k_B T << T_s$) is given by

$$G_t(T) = G_e[1+\sqrt{\frac{2}{\pi}}\frac{k_B T}{T_s} - \frac{2}{\sqrt{\pi}}(\frac{k_B T}{T_s})^3] \qquad (2)$$

where $G_e$ is the temperature independent average conductance of electrons. In addition to this, the thermal excitation of the carriers can also contribute to temperature dependent conductivity. Thus the total conductance at CNP at low temperatures ($k_B T << T_s$) is given by:

$$G_t(T) = G_e[1+\sqrt{\frac{2}{\pi}}\frac{k_B T}{T_s} + \frac{\pi^2}{6}(\frac{k_B T}{T_s})^2] \qquad (3)$$

At high temperatures ($k_B T >> T_s$), conductivity due to electron-hole puddles approaches a limit and conductivity arising from the thermal excitation of carriers through the change in carrier density is given by:

$$G_t(T) = G_e[2-\sqrt{\frac{2}{\pi}}\frac{T_s}{k_B T} + \frac{1}{2}(\frac{T_s}{k_B T})^2] \qquad (4)$$

We observe that the above equations do not give a good fit to our experimental results as seen from Fig. 5(a). Note that the corresponding temperature of $T_s$ at CNP is 173 K.

Figure 5(b) shows the logarithmic dependence of the conductance $G/G_{5K}$ as a function of the inverse of the temperature at CNP. If there is only an activated process, one should have a single slope in the entire temperature range of the measurement. The conductance in the high and intermediate temperatures is activated, whereas at low temperatures it does not have any activation. Based on the report of variable range hoping (VRH) transport in graphitic nanocrystals[29] and epitaxial graphene,[30] we plot a graph of $G/G_{5K}$ versus $T^{(-1/3)}$ in the inset of Fig. 5(b) and the absence of a single slope suggests that two dimensional VRH mechanism cannot explain the transport behavior in our case. Therefore, we conclude that our proposed empirical formula of Eq. (1) correctly describes the transport properties of gated bilayer graphene.

In conclusion, the study of temperature dependent electrical transport properties of gated bilayer graphene devices reveals a clear evidence of insulating behavior due to electron-hole charge puddles. The electrical resistivity increases while the mobility decreases with decreasing the temperature, a characteristic of carrier inhomogeneous graphene. The theoretical fittings using a formula from single electron tunneling indicate that electrical resistivity follows a universal curve with a scaling parameter. The scaling parameter is determined to be a measure of the electron-hole puddle correlation. We rule out a possibility of the Kondo effect due to a positive magnetoresistance in our samples.

This work is supported by NRF-CRP “Novel 2D materials with tailored properties: beyond graphene” (No. R-144-000-295-281).

**Figure Captions**

FIG. 1. (a) Raman spectrum of bilayer graphene. (b) 2D peak with the theoretical fit.

FIG. 2. (a) Optical micrograph of the patterned top gated graphene device. "S" stands for source, "G" for top gate, and "D" for drain. (b) Resistance with respect to that at 5 K ($R_{xx}/R_{5K}$) versus top gate voltage ($V_G$) at different temperatures from 5 to 220 K. (c) Optical micrograph of the patterned Hall bar graphene device with a back gate. (d) Four probe resistivity (ρ) versus back gate voltage ($V_{BG}$) at different temperatures from 60 to 220 K.

FIG. 3. (a) Resistance ($R_{xx}/R_{5K}$) versus temperature ($T$) as a function of top gate voltage ($V_G$) with the fits. (b) Four probe resistivity (ρ) versus temperature ($T$) as a function of back gate voltage ($V_{BG}$) with the fits. (c) Normalized resistance versus normalized temperature at different $V_G$ showing the universal nature of the scattering in bilayer graphene. (d) Normalized resistance versus normalized temperature at different $V_{BG}$.

FIG. 4. (a) Calculated scaling parameter ($T_s$) versus top gate voltage ($V_G$). (b) Normalized residual resistivity ($R_0/R_{max}$) versus $V_G$. (c) Normalized saturated resistivity ($R_e/R_{max}$) versus $V_G$. (d) Calculated conductance ($G_0$) versus $V_G$.

FIG. 5. (a) $G(T)/G_{5K}$ versus temperature ($T$) with theoretical fits using Eq. (2-4). (b) Logarithmic dependence of $G(T)/G_{5K}$ as a function of $1000/T$ at $V_G = 0.5$ V. The absence of a single slope rules out a simple semiconducting behavior. The inset shows the logarithmic dependence of $G(T)/G_{5K}$ as a function of $T^{(-1/3)}$. The absence of a single slope rules out variable range hoping (VRH) mechanism.

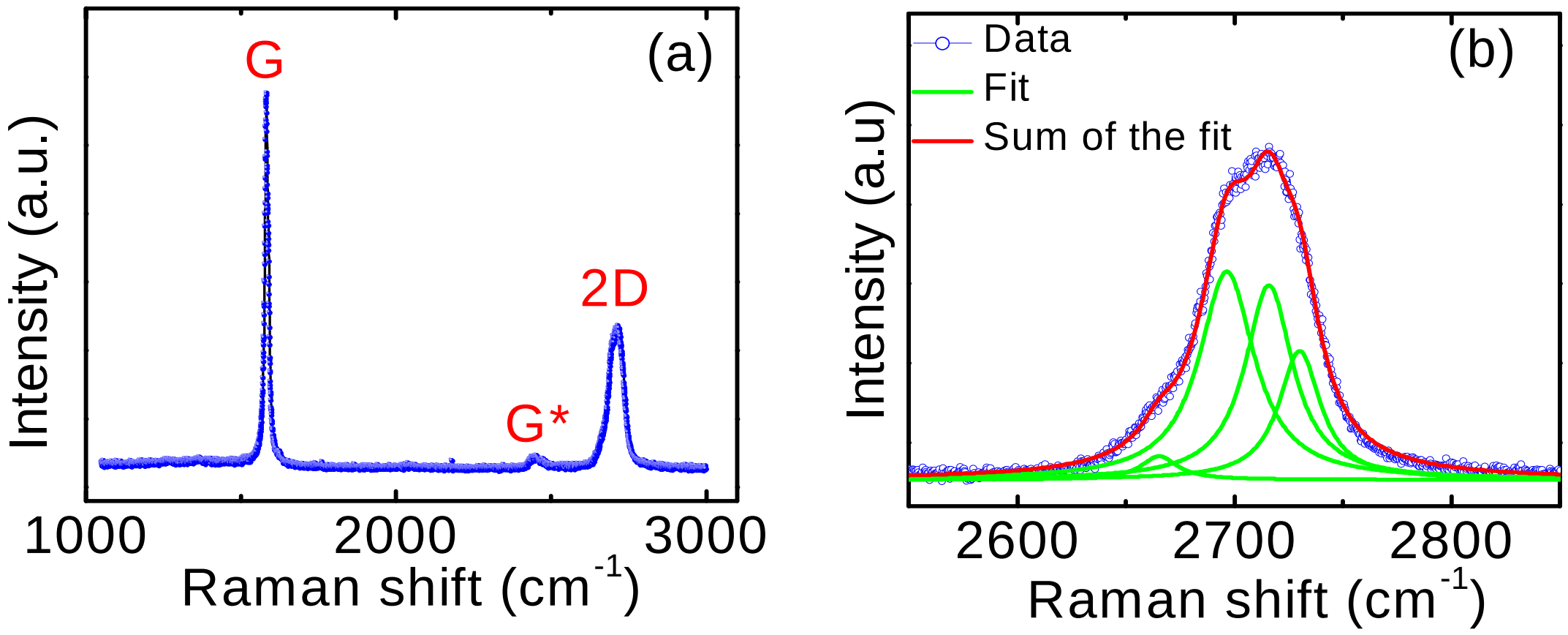

Figure 1

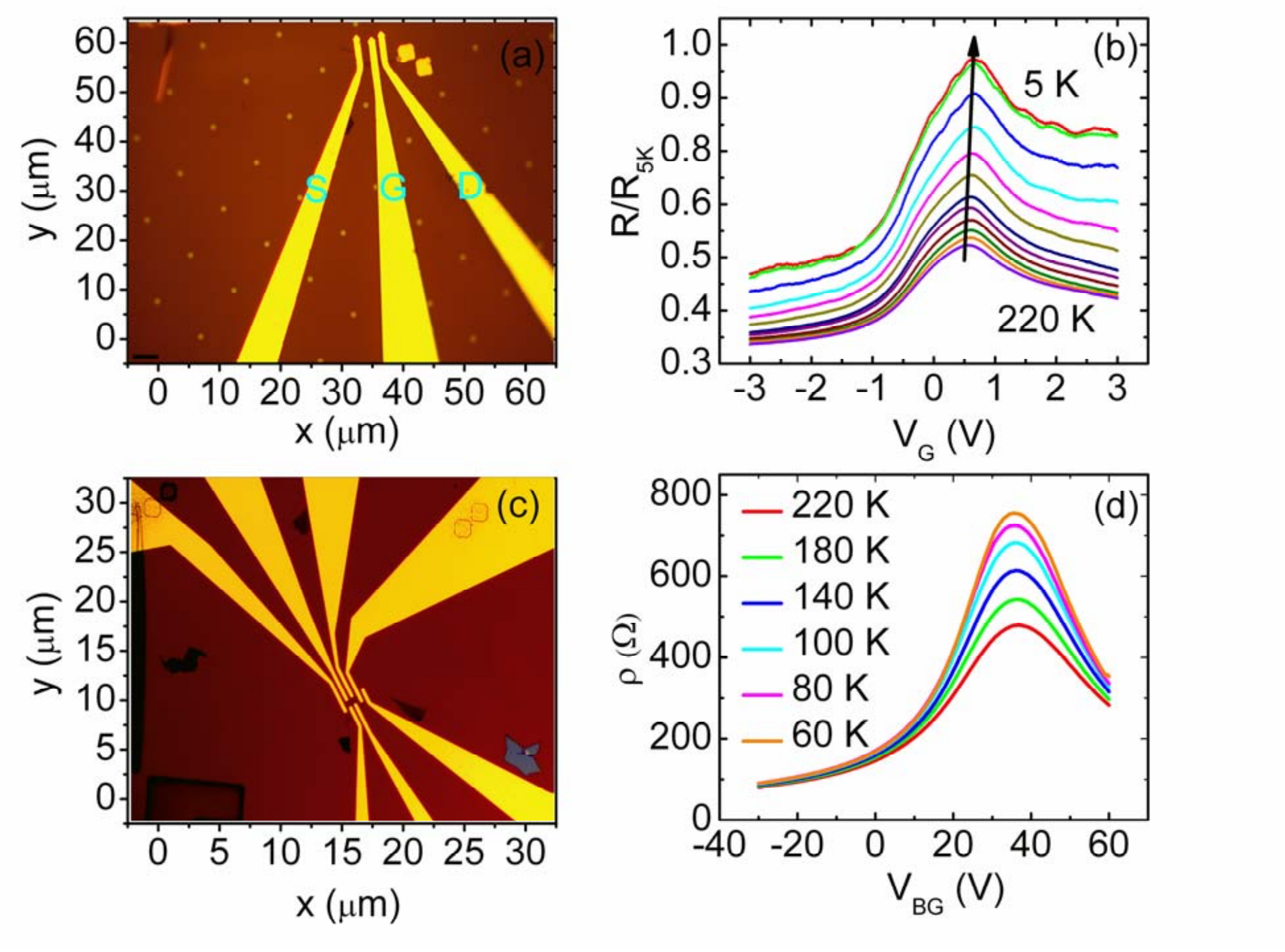

Figure 2

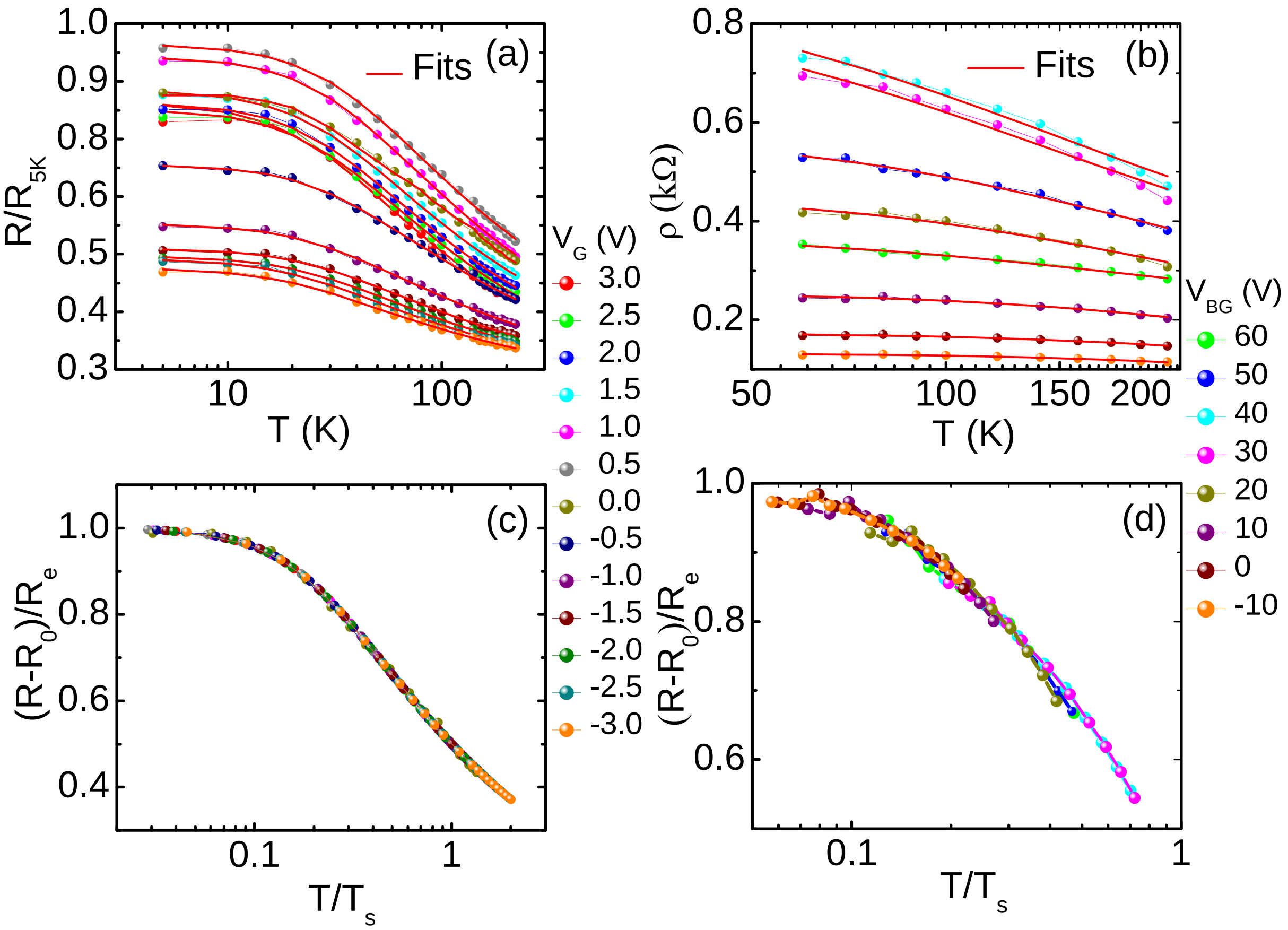

Figure 3

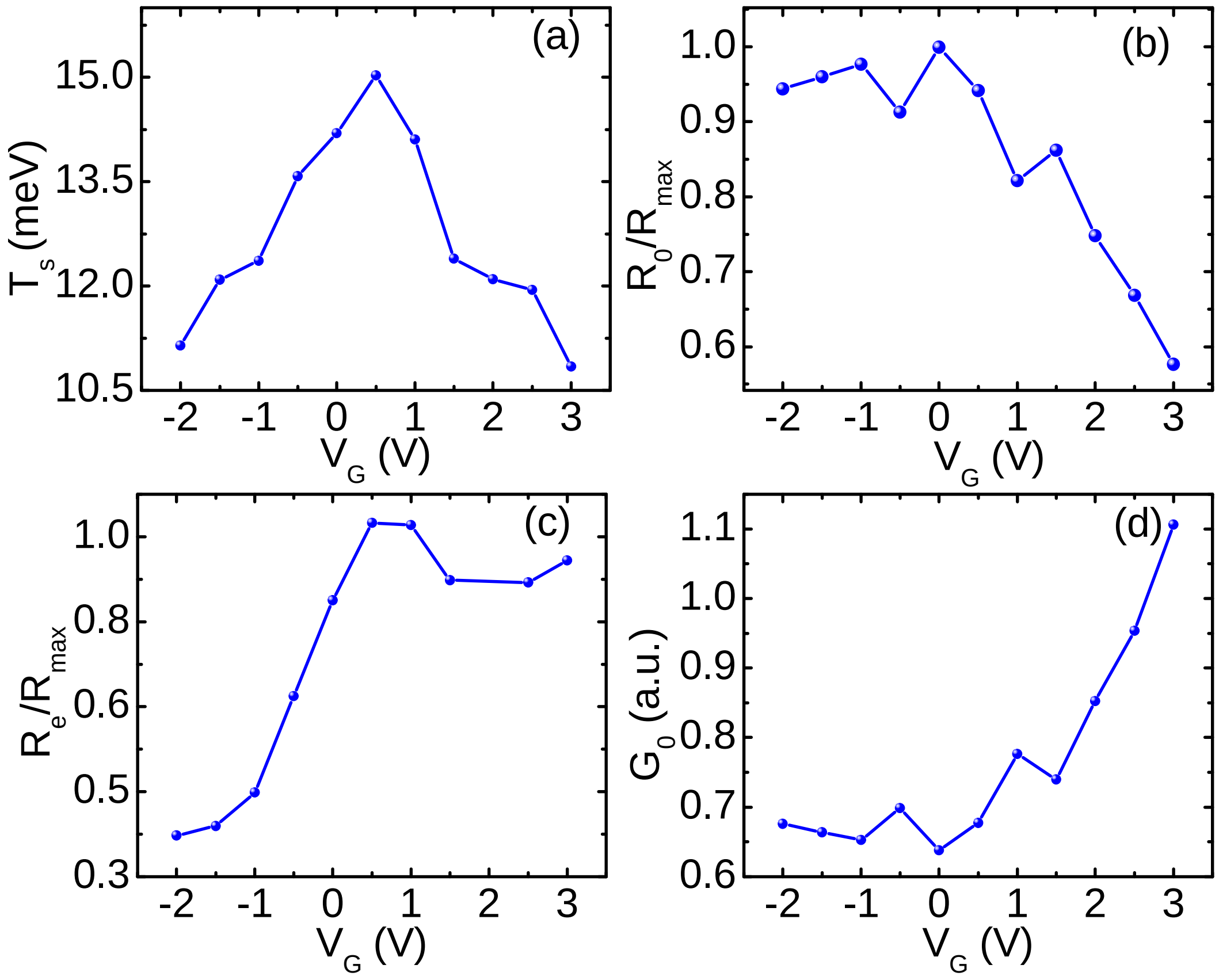

Figure 4

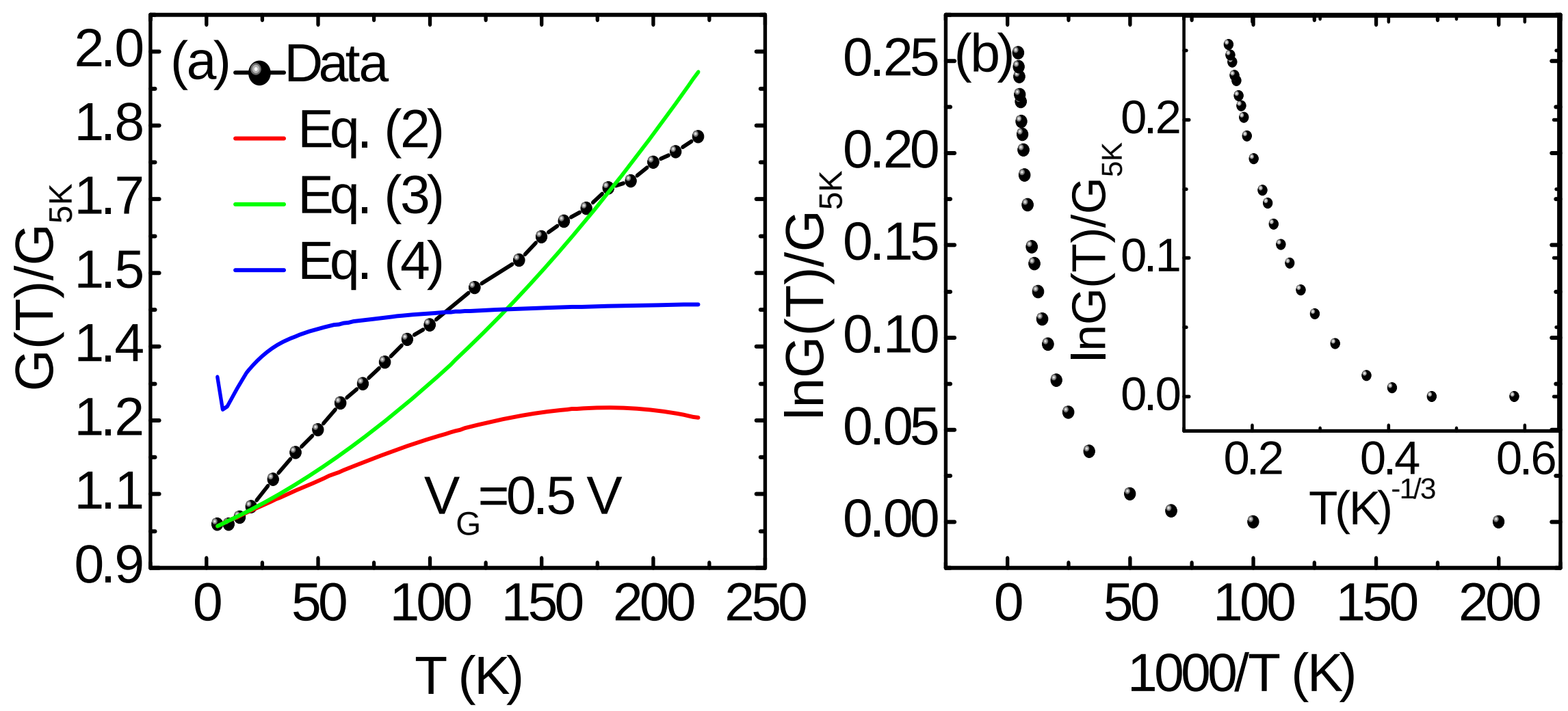

Figure 5